\documentstyle[preprint,12pt,aps,tighten,eqsecnum,floats]{revtex}

\newcommand{\beq}{\begin{equation}}
\newcommand{\eeq}{\end{equation}}
\newcommand{\beqa}{\begin{eqnarray}}
\newcommand{\eeqa}{\end{eqnarray}}

\renewcommand{\Re}{\mbox{Re}}
\renewcommand{\Im}{\mbox{Im}}

\def\lsim{\mathrel{\rlap{\lower4pt\hbox{\hskip.1pt$\sim$}}
    \raise1pt\hbox{$<$}}}         
\def\gsim{\mathrel{\rlap{\lower4pt\hbox{\hskip.1pt$\sim$}}
    \raise1pt\hbox{$>$}}}         

\begin{document}
\preprint{
\vbox{  \hbox{WIS/25/01-Nov-DPP}
	\hbox{LBNL-49160}
        \hbox{hep-ph/0202117}
        \hbox{} }}

\title{New Physics and Future B Factories}
\author{Zoltan Ligeti$^{\,1\,}$\thanks{zligeti@lbl.gov} and
Yosef Nir$^{\,2\,}$\thanks{yosef.nir@weizmann.ac.il}}
\vskip 1cm
\address{$^1$Ernest Orlando Lawrence Berkeley National Laboratory \\
University of California, Berkeley, CA 94720, USA \\
$^2$ Department of Particle Physics, Weizmann Institute of Science\\
 Rehovot 76100, Israel}
\maketitle
\thispagestyle{empty}
\setcounter{page}{0}
\begin{abstract}%
Further experimental and theoretical studies of the physics of flavor
and CP violation are well motivated. Within the supersymmetric framework,
higher precision measurements will allow to explore classes of models with
stronger degree of universality: first, models with no universality, such
as alignment or heavy first two squark generations; second, models with
approximate universality, such as dilaton dominance or AMSB; and finally
models of exact universality, such as GMSB. A broad program, including
various rare processes or CP asymmetries in $B$, $D$ and $K$ decays,
will provide detailed information about viable extensions of the
Standard Model. Some highlights of future $B$-physics experiments
(the present $B$-factories with integrated luminosity of 0.5 ab$^{-1}$,
hadron machines, and future high-luminosity $B$-factories) are described.

\end{abstract}

\bigskip\bigskip\bigskip

\centerline{\it Lecture given at the}
\centerline{\bf Fifth KEK Topical Conference}
\centerline{\bf Frontiers in Flavor Physics}
\centerline{\it KEK, Tsukuba, Japan}
\centerline{\it November 20 $-$ 22 2001}

\vfill

\section{Introduction}
All existing measurements of flavor and CP violation are consistent with
the CKM framework. In particular, the two recent measurements of CP violation in
$B$ decays \cite{Aubert:2001nu,Abe:2001xe} have provided the first precision
test of CP violation in the Standard Model. Since the model has passed this
test successfully, we are able, for the first time, to make the following
statement: {\it The Kobayashi-Maskawa phase is, very likely,
the dominant source of CP violation in low-energy flavor-changing processes.}

Still, further experimental and theoretical investigations of flavor and CP
violation are well motivated. Here are some of the reasons for the interest
in these aspects of high energy physics:

(i) {\it The flavor puzzle.}
The flavor parameters of the Standard Model, that is the fermion masses
and mixing angles, are small (except for $m_t$ and $\delta_{\rm KM}$) and
hierarchical. The Standard Model offers no explanation for these puzzling
features. Perhaps the special structure of the Yukawa couplings is a hint
of new physics.

(ii) {\it The supersymmetric flavor puzzle.}
Flavor changing neutral current processes are highly suppressed in Nature.
This experimental fact is nicely accounted for within the Standard Model, where
flavor changing neutral currents are absent at tree level. They appear at the
loop level, but then they are suppressed very effectively by the small CKM
angles and the GIM mechanism. Various extensions of the Standard Model need to
have very special flavor structures in order to achieve similarly effective
suppression mechanisms. The most striking example is supersymmetry. While
tree level FCNC can be forbidden (together with baryon and lepton number
violations) by imposing an $R_p$ symmetry, there is no similarly natural and
generic way to suppress the loop contributions to flavor and CP violating
processes. Experimental studies of flavor physics are then crucial for
understanding the mechanism of dynamical supersymmetry breaking.

(iii) {\it New sources of CP violation.}
Almost any extension of the Standard Model provides new sources of CP
violation. These sources often allow for significant deviations from the
Standard Model predictions. Moreover, various CP violating observables can be
calculated with very small hadronic uncertainties. Consequently, CP violation
provides an excellent probe of new physics.

(iv) {\it The strong CP problem.}
It is presently not understood why CP violation is so small in the strong
interactions. The upper bound on the electric dipole moment of the neutron
constrains nonperturbative CP violating QCD effects to be at least ten orders
of magnitudes below naive expectations.

(v) {\it Baryogenesis.}
The observed baryon asymmetry of the universe requires that CP is violated,
but quantitatively it cannot be accounted for by the Kobayashi-Maskawa
mechanism. It is clear then that new sources of CP violation must exist
in Nature.

The experimental program of $B$ physics has just entered a new era
in precision, in sensitivity and in probing time-dependent CP asymmetries.
We are trying to overconstrain the CKM parameters, and to test the Standard
Model correlations and (approximate) zeros. Experimental studies of $B$ decays
are guaranteed to enrich our understanding of flavor and CP physics:

(i) At the very least, these experiments will significantly improve the
determination of the CKM parameters.

(ii) If low-energy supersymmetry is realized in Nature then, as explained
above, the study of flavor physics (whether consistent or inconsistent with the
Standard Model predictions) will provide unique information about high scale
physics. We explain the relations between the mechanism of dynamical
supersymmetry breaking and the flavor and CP physics in the next section.

(iii) At best, measurements of rare $B$ decays and related CP violation will
allow us to make progress on the road to solving the flavor puzzle and/or the
puzzle of baryogenesis.
It is important, however, to realize that, in contrast to the fine-tuning
problem of electroweak symmetry breaking, here there is no analogous argument
that says that the relevant new physics must appear at a scale that is not too
far above $m_Z$. The scale of new flavor or CP violating physics may
be very high, well beyond the reach of $B$ factories.

\section{The Supersymmetric Framework}
Supersymmetry solves the fine-tuning problem of the Standard Model and has
many other virtues. But at the same time, it leads to new problems:
baryon number violation, lepton number violation, large flavor changing
neutral current processes and large CP violation. The first two problems
can be solved by imposing $R$-parity on supersymmetric models. There is no
such simple, symmetry-related solution to the problems of flavor and CP
violation. Instead, suppression of the relevant couplings can be achieved
by demanding very constrained structures of the soft supersymmetry breaking
terms. There are two important questions here: First, can theories of dynamical
supersymmetry breaking (for a review, see \cite{Shadmi:2000jy}) naturally
induce such structures?
Second, can measurements of flavor changing and/or CP violating processes
shed light on the structure of the soft supersymmetry breaking terms?
Since the answer to both questions is in the affirmative, we conclude that
flavor changing neutral current processes and CP violating observables will
provide clues to the crucial question of how supersymmetry breaks.

\subsection{Flavor and CP Problems}
A generic supersymmetric extension of the Standard Model contains a host of new
flavor and CP violating parameters. (For reviews of CP violation in
supersymmetry see \cite{Grossman:1997pa,Dine:2001ne}.) In fact, the Lagrangian
of the minimal supersymmetric Standard Model has 124 physical parameters: 80
real ones and 44 imaginary ones \cite{Haber:1998if}. Most of these parameters
are related to flavor changing couplings. In addition to the Yukawa terms of
the Standard Model, we now have flavor violation in trilinear scalar
couplings ($A$-terms) and scalar mass-squared matrices ($\tilde m^2$-matrices).
In contrast to the SM, we now have also flavor diagonal phases, coming from
the bilinear Higgsino coupling (the $\mu$-term), the bilinear Higgs coupling
(the $B$-term) and gaugino masses ($m_{\tilde g_a}$). Supersymmetry provides
an impressive demonstration that low energy flavor physics might be richer than
the CKM framework.

The requirement of consistency with experimental data provides strong
constraints on many of these parameters. For this reason, the physics of flavor
and CP violation has had a profound impact on supersymmetric model building.
The supersymmetric flavor problem and the supersymmetric CP problem are
well represented by the predictions for the mass difference ($\Delta m_K$)
and CP violation ($\varepsilon_K$) in $K^0-\overline{K^0}$ mixing and
by the electric dipole moment of the neutron ($d_N$).

The supersymmetric contribution to $\Delta m_K$ is dominated
by diagrams involving $Q$ ($SU(2)_{\rm L}$ doublet) and $\bar D$
(down singlet) squarks in the same loop. To simplify our presentation,
we assume that there is a single scale $\tilde m$ that characterizes
all supersymmetry breaking terms, that is, $\tilde m\simeq m_{\tilde g}\simeq
m_{\tilde Q} \simeq m_{\tilde D}$ (our results depend only weakly on this
assumption). We focus on the contribution from the first two squark
families (see, for example, \cite{Gabbiani:1996hi}):
\beq\label{epsKScon}
{(\Delta m_K)^{\rm SUSY}\over(\Delta m_K)^{\rm EXP}}
\sim10^5\left ({300 \ GeV\over\tilde m}\right)^2
\left({m^2_{\tilde Q_2}-m^2_{\tilde Q_1}\over \tilde m^2}\right)
\left({m^2_{\tilde D_2}-m^2_{\tilde D_1}\over \tilde m^2}\right)
\Re\left[(K^d_L)_{12}(K^d_R)_{12}\right],
\eeq
where $K^d_L$ ($K^d_R$) are the mixing matrices in the gluino
couplings to left-handed (right-handed) down quarks and their scalar partners.
The constraint from $\varepsilon_K$ can be obtained by replacing
$10^5\Re\left[(K^d_L)_{12}(K^d_R)_{12}\right]$ with
$10^7\Im\left[(K^d_L)_{12}(K^d_R)_{12}\right]$.
In a generic supersymmetric framework, we expect $\tilde m={\cal O}(m_Z)$,
$\Delta m_{\tilde Q,\tilde D}^2/\tilde m^2={\cal O}(0.1)$ and
$(K^d_{L,R})_{ij}={\cal O}(1)$.
(The approximate degenracy in squark masses is induced by RGE if the soft
breaking terms are all induced close to the Planck scale with comparable size.)
Then the constraint (\ref{epsKScon}) is generically violated by about
three orders of magnitude. Eq. (\ref{epsKScon}) also suggests three
possible ways to solve the supersymmetric flavor problems:

(i) Heavy squarks: $\tilde m\gg300\ GeV$;

(ii) Universality: $\Delta m_{\tilde Q,\tilde D}^2\ll \tilde m^2$;

(iii) Alignment: $|(K^d_M)_{12}|\ll1$;

In addition, the related CP problems are alleviated if the relevant phases
fulfill $\sin\phi\ll1$.

Supersymmetry predicts also flavor preserving CP violation.
For simplicity, we describe this aspect in a supersymmetric model without
additional flavor mixings, {\it i.e.} the minimal supersymmetric standard model
(MSSM) with universal sfermion masses and with the $A$-terms proportional to
the corresponding Yukawa couplings.  In such a constrained framework, there are
two new physical phases \cite{Dugan:1985qf,Dimopoulos:1996kn}: $\phi_A$, which
is related to the relative phase between the $A$-terms and the gaugino masses,
and $\phi_B$, which is related to the relative phase between the $\mu$-term and
the $B$-term. The most significant effect of $\phi_A$ and $\phi_B$ is their
contribution to electric dipole moments (EDMs). For $d_N$, we obtain (see,
for example, \cite{Fischler:1992ha}):
\beq\label{dipole}
{d_N\over6.3\times10^{-26}\ e\, {\rm cm}}
\sim 300\, \left({100\, GeV\over \tilde m}\right )^2\sin \phi_{A,B},
\eeq
where the denominator on the left hand side gives the present experimental
upper bound. In a generic supersymmetric framework, we expect
$\tilde m={\cal O}(m_Z)$ and $\sin\phi_{A,B}={\cal O}(1)$. Then the experimental
bound is generically violated by about two orders of magnitude. This is {\it
the Supersymmetric CP Problem}. Eq. (\ref{dipole}) shows two possible ways to
solve the supersymmetric CP problem:

(i) Heavy squarks: $\tilde m\gsim1\ TeV$;

(ii) Approximate CP: $\sin\phi_{A,B}\ll1$

\subsection{Supersymmetry Breaking and Universality}
Two scales play an important role in our discussion of supersymmetry:
$\Lambda_S$, where the soft supersymmetry breaking terms are generated, and
$\Lambda_F$, where flavor dynamics takes place. When $\Lambda_F\gg\Lambda_S$,
it is possible that there are no genuinely new sources of flavor and CP
violation. This leads to models with exact universality. When
$\Lambda_F\lsim\Lambda_S$, we do not expect, in general,
that flavor and CP violation are limited to the Yukawa matrices.
This leads to models without universality. In some special cases of
supersymmetry breaking with $\Lambda_F\lsim\Lambda_S$, it is possible that the
leading contributions to supersymmetry breaking are universal. But it is
difficult to avoid subdominant flavor-dependent contributions. Then we
expect approximate universality.

\subsubsection{Exact Universality: Gauge Mediated Supersymmetry Breaking}
If at some high energy scale squarks are exactly degenerate and the $A$ terms
proportional to the Yukawa couplings, then the contributions to FCNC come from
RGE and are GIM suppressed, for example,
\beq\label{gimsup}
\Delta m_K\propto\Re[(V_{td}V_{ts}^*)^2]Y_t^4
\left[{\log(\Lambda_S/m_W)\over16\pi^2}\right]^2.
\eeq
This contribution is negligibly small.

In models of Gauge Mediated Supersymmetry Breaking (GMSB)
\cite{Dine:1995vc,Dine:1996ag,Dine:1997xk}, superpartner masses are generated
by the SM gauge interactions. These masses are then exactly universal at the
scale $\Lambda_S$ at which they are generated  (up to tiny high order effects
associated with Yukawa couplings). Furthermore, $A$ terms are suppressed by
loop factors. The only contribution to FCNC is then from the running, and since
$\Lambda_S$ is low it is highly suppressed. Similarly to $K-\bar K$ mixing, the
supersymmetric contribution to $D-\bar D$ mixing is small and we expect no
observable effects. Supersymmetric contributions to  $B-\bar B$ mixing are,
at most, 20\% of the SM one and usually much smaller. Such deviations are too
small to be signalled by $\Delta m_B$ but can perhaps give observable effects
in the CP asymmetries in $B\to\psi K$ decays.

More generally, in any supersymmetric model where there are no new flavor
violating sources beyond the Yukawa couplings, FCNC and CP violation in meson
decays are hardly modified from the SM predictions \cite{Demir:2000qm}.

\subsubsection{Approximate Universality: Gravity, Anomaly and Gaugino Mediation}
If different moduli of string theory obtain supersymmetry breaking $F$ terms,
they would typically induce flavor-dependent soft terms through their
tree-level couplings to Standard Model fields. There are however various
scenarios in which the leading contribution to the soft terms is flavor
independent. The three most intensively studied frameworks are dilaton
dominance, anomaly mediation and gaugino mediation.

{\bf Dilaton dominance} assumes that the dilaton $F$ term is the dominant one.
Then, at tree level, the resulting soft masses are universal and the $A$ terms
proportional to the Yukawa couplings \cite{Barbieri:1993jk}. Both universality
and proportionality are, however, violated by string loop effects. These induce
corrections to squark masses of order ${\alpha_X\over\pi}m^2_{3/2}$, where
$\alpha_X=[2\pi(S+S^*)]^{-1}$ is the string coupling. There is no reason why
these corrections would be flavor blind. The effect of these terms is, however,
somewhat suppressed by RGE effects which enhance the universal part of the
squark masses by roughly a factor of five, while leaving the off-diagonal
entries essentially unchanged. The flavor suppression factor is
then \cite{Louis:1995ht}
\beq\label{dildoe}
{\Delta m^2_{12}\over\tilde m^2}\simeq{m^{2\ {\rm one-loop}}_{12}\over
m^2_{\tilde g}}\simeq{\alpha_X\over\pi}{1\over25}\simeq4\times10^{-4}\ .
\eeq
Dilaton dominance relies on the assumption that loop corrections are small.
This probably presents the most serious theoretical difficulty for this idea,
because it is hard to see how non-perturbative effects, which are probably
required to stabilize the dilaton, could do so in a region of weak coupling.
In the strong coupling regime, these corrections could be much larger.
However, this idea at least gives some plausible theoretical explanation
for how universal masses might emerge in hidden sector models. Given that
dilaton stabilization might require that non-perturbative effects are
important, the flavor suppression might in reality be weaker than the
estimate of eq.~(\ref{dildoe}).

{\bf Anomaly mediation} (AMSB) provides another approach to solving the flavor
problems of supersymmetric theories, as well as to obtaining a predictive
spectrum. The conformal anomaly of the Standard Model gives rise to soft
supersymmetry breaking terms for the Standard Model fields
\cite{Randall:1999uk,Giudice:1998xp}. These terms are generated purely by
gravitational effects and are universal. In general, naturalness considerations
suggest that couplings of hidden and visible sectors should appear in the
Kahler potential, leading to soft masses for scalars already at tree level, and
certainly by one loop. As a result, one would expect the anomaly-mediated
contributions to be irrelevant. However, in ``sequestered sector models''
\cite{Randall:1999uk}, in which the visible sector fields and supersymmetry
breaking fields live on different branes separated by some distance, the
anomaly mediated contribution could be the dominant effect. This leads to a
predictive picture with universal scalar masses. It has been realized, however,
that within the framework of string/M theories, the separation of branes is, by
itself, not enough to avoid tree-level, non-universal squark and slepton masses
\cite{Bagger:2000rd,Anisimov:2001zz,Luty:2001zv}. Only under special
conditions, such as compactification to pure five-dimensional supergravity
with end of the world branes, is the anomaly-mediated contribution dominant.
Quite generically, however, sub-dominant effects give deviations from
universality, {\it e.g.} \cite{Anisimov:2001zz}
\beq\label{amomed}
{\Delta m^2_{12}\over\tilde m^2}={\cal O}\left({T\over S}\right),
\eeq
where $S$ is the dilaton and $T$ is some modulus. On both theoretical and
phenomenological grounds, one expects that the ratio $T/S$ is not much smaller
than unity, perhaps $T/S\sim1/3$. Such non-universal contributions may easily
violate the $\Delta m_K$ and $\varepsilon_K$ constraints \cite{Dine:2001ne}.
(Another difficulty of this framework is related to the fact that slepton
masses-squared are negative, so modification is required.) Similar comments
apply to the framework of {\bf gaugino mediation} ($\tilde g$MSB)
\cite{Kaplan:2000ac,Chacko:2000mi}. These models also
suppress dangerous tree level contact terms by invoking extra dimensions, with
the Standard Model matter fields localized on one brane and the supersymmetry
breaking sector on another brane. In this case, however, the Standard Model
gauge fields are in the bulk, so gauginos get masses at tree level, and as a
result scalar masses are generated by running. Again, however, non-universal
tree and one loop contributions to scalar masses are generic and
significant violations of degeneracy and proportionality are expected.

\subsubsection{No Universality: Supersymmetric Horizontal Symmetries}
Various frameworks have been suggested in which flavor symmetries,
designed to explain the hierarchy of the Yukawa couplings, impose at the
same time a special flavor structure on the soft supersymmetry breaking
terms that helps to alleviate the flavor and CP problems.

{\bf Alignment} models do not assume any squark degeneracy.
Instead, flavor violation is suppressed because the squark mass matrices are
approximately diagonal in the quark mass basis. This is the case in models of
Abelian flavor symmetries, in which the off-diagonal entries in both the quark
mass matrices and in the squark mass matrices are suppressed by some power of a
small parameter, $\lambda$, that quantifies the breaking of some Abelian flavor
symmetry. A natural choice for the value of $\lambda$ is $\sin\theta_C$, so we
will take $\lambda\sim0.2$. One would naively expect the first two generation
squark mixing to be of the order of $\lambda$. However, the $\Delta m_K$
constraint is not satisfied with the `naive alignment', $K_{12}^d\sim\lambda$,
and one has to construct more complicated models to achieve the required
suppression \cite{Nir:1993mx,Leurer:1994gy}.
As concerns $D-\bar D$ mixing, models of alignment are very predictive.
It is unavoidable in this framework that, to a very good approximation,
\beq\label{mixali}
|(K^u_L)_{12}|=\sin\theta_C.
\eeq
Consequently, the supersymmetric contributions to $\Delta m_D$ are close
to the present experimental bound. Furthermore, there is no reason for
the related CP violating phase to be small. Thus, large CP violating effects
could be observed in the doubly-Cabibbo-suppressed $D\to K\pi$ decays.
The effects in $B-\bar B$ mixing are smaller: one expects $|(K^d_L)_{13}|\sim
|V_{ub}|$, leading to a few percent effects on CP asymmetries in neutral
$B$ decays.
We conclude that one can construct models in which an Abelian horizontal
symmetry solves the supersymmetric problems of flavor and CP. These models
are however not the generic ones in this framework. They can be tested
through measurements of mixing and CP violation in the neutral $D$ system
and searched for through small but perhaps non-negligible effects in
CP violation in neutral $B$ decays.

{\bf Non-Abelian horizontal symmetries} can induce approximate degeneracy
between the first two squark generations, thus relaxing the flavor and CP
problems \cite{Dine:1993np}. (A review of $\varepsilon_K$ in this class of
models can be found in \cite{Grossman:1997pa}.) The approximate degeneracy
between the first two squark generations suppresses also the supersymmetric
contributions to $D-\bar D$ mixing. Small but perhaps observable deviations
from the Standard Model predictions for CP asymmetries in $B$ decays are
possible. Similar to models of Abelian flavor symmetries, one can construct
models of non-Abelian symmetries in which the symmetry solves both the
$\varepsilon_K$ and the $d_N$ problems. These models are however not the
generic ones in this framework.

Finally, one can construct models of {\bf heavy first two generation squarks}.
Here, the basic mechanism to suppress flavor changing processes is actually
flavor diagonal: $m_{\tilde q_{1,2}}\sim20\ {\rm TeV}$. Naturalness does not
allow higher masses, but this mass scale is not enough to satisfy even the
$\Delta m_K$ constraint \cite{Cohen:1997sq}, and one has to invoke alignment,
$K_{12}^d\sim\lambda$. This is still not enough to satisfy the $\varepsilon_K$
constraint, and a somewhat small phase is required.
Two more comments are in order: First, in this framework, gauginos are
significantly lighter than the first two generation squarks, and so RGE cannot
induce degeneracy. Second, the large mass of the squarks is enough to solve the
EDM related problems, and so it is only the $\varepsilon_K$ constraint that
motivates a special phase structure.

\subsection{Flavor and CP Violation as a Probe of Supersymmetry}
We have seen that supersymmetric flavor models can be roughly divided
to three classes:

(i) Models of exact universality, where the only effects of flavor violation
and CP violation come from the Yukawa sector and enter through RGE.

(ii) Models of approximate universality, where there are genuinely new
sources of flavor and CP violation which are, however, subleading to the
dominant universal structure.

(iii) Models without universality, where horizontal symmetries (or large
masses) suppress flavor and CP violation.

The latter class is the easiest to explore through flavor and CP violation,
since it gives the largest effects. With more precise measurements we can
probe stronger degrees of universality.

We would like to emphasize the following points:

(i) For supersymmetry to be established, a direct observation of supersymmetric
particles is necessary. Once it is discovered, then measurements of CP
violating observables will be a very sensitive probe of its flavor structure
and, consequently, of the mechanism of dynamical supersymmetry breaking.

(ii) It seems possible to distinguish between models of exact universality and
models with genuine supersymmetric flavor and CP violation. The former tend to
give $d_N\lsim10^{-31}$ e cm while the latter usually predict
$d_N\gsim10^{-28}$ e cm.

(iii) The proximity of $a_{\psi K}$ to the SM predictions is obviously
consistent with models of exact universality. It disfavors models of heavy
squarks such as that of ref. \cite{Cohen:1997sq}. Models of flavor symmetries
allow deviations of order 20\% (or smaller) from the SM predictions. For such
new physics to be convincingly established, the hadronic uncertainties that
affect the SM allowed range of $a_{\psi K}$ will be required to be reduced
well below this level \cite{Eyal:2000ys}.

(iv) Alternatively, the fact that $K\to\pi\nu\bar\nu$ decays are not affected
by most supersymmetric flavor models
\cite{Nir:1998tf,Buras:1998ij,Colangelo:1998pm}
is an advantage here. The Standard Model correlation between
$a_{\pi\nu\bar\nu}$ and $a_{\psi K}$ is a much cleaner test than
a comparison of $a_{\psi K}$ to the CKM constraints.

(v) The neutral $D$ system provides a stringent test of alignment.
Observation of CP violation in the $D\to K\pi$ decays will make a convincing
case for new physics.

\section{Highlights of Higher Luminosity B Factories}
Considering possible outcomes of future measurements, there are different
time-scales to look at.  Until 2005 or so, BABAR \cite{Harrison:1998yr} and
BELLE expect to collect $\sim 0.5\,\mbox{ab}^{-1}$ of data each. On this
timescale the Tevatron \cite{Anikeev:2001rk} will also yield important
information, most crucially $B_s$ mixing.  In the second half of the decade
dedicated hadronic $b$ factories will start to operate, and the $e^+e^-$
machines may undergo upgrades to luminosities in the $10^{35}-10^{36}\,
\mbox{cm}^{-2}\, \mbox{sec}^{-1}$ range~\cite{babar36,belle35}.

We remind the reader that the cross section of $B$ production in the
$e^+e^-$ machines is 1.2 nb, so that a luminosity of $3\times10^{33}\,
\mbox{cm}^{-2}\, \mbox{sec}^{-1}$ gives $1.8\times10^7$ $B^0\overline{B^0}$
pairs (and a similar number of $B^+B^-$ pairs) in a year. For comparison, the
corresponding cross section in BTeV is about 100 $\mu$b, so that a luminosity
of $2\times10^{32}\, \mbox{cm}^{-2}\, \mbox{sec}^{-1}$ gives $2\times10^{11}$
$b\overline{b}$ pairs. In LHC-B the cross section is even larger.

\subsection{\boldmath $B_s$ mixing and $|V_{ub}|$}
In the Standard Model these measurements will determine the two sides of the
unitarity triangle.   Soon after $B_s$ mixing is observed, the experimental
error on $\Delta m_d/\Delta m_s$ is expected to be reduced below the 1\%
level.  Thus, the uncertainty of  $|V_{td}/V_{ts}|$  will be dominated by the
error of $(f_{B_d}/f_{B_s})\sqrt{B_{B_d}/B_{B_s}}$ from lattice QCD.  This is
presently at the $4-5\%$ level in unquenched calculations with two  light quark
flavors~\cite{JLQCD}, and it is important to reduce this  source of uncertainty
using simulations with three light flavors.

For the determination of $|V_{ub}|$ from inclusive semileptonic $B$ decays, the
reconstruction of $q^2$ and $m_X$ (lepton--neutrino invariant mass and hadronic
invariant mass) offers probably the smallest theoretical error~\cite{bll}.  The
error in determining $|V_{ub}|$ from exclusive semileptonic $B$ decays will be
controlled by the accuracy of unquenched lattice calculations.  With
$0.5\,\mbox{ab}^{-1}$, a determination of $|V_{ub}|$ at the $5-10\%$ level
should be possible.  Confidence in such a precision will come from consistency
between different model independent determinations.

\subsection{\boldmath $a_{\pi\pi}$ and $a_{\rho\pi}$}
Measuring the CP asymmetries in $b\to u\bar ud$ transitions will add
a significant constraint on the unitarity triangle. The simplest process
involving a final CP eigenstate is $B\to\pi^+\pi^-$. The problem here is that
the penguin contribution is nonnegligible compared to the tree contribution,
$r^{\pi\pi}_{PT}\equiv P_{\pi\pi}/T_{\pi\pi}={\cal O}(0.3)$, where $P_{\pi\pi}$
and $T_{\pi\pi}$ are defined through the CKM decomposition,
\beq\label{abpipi}
A(B^0\to\pi^+\pi^-)=T_{\pi\pi}V_{ub}^*V_{ud}+P_{\pi\pi}V_{tb}^*V_{td}.
\eeq
One needs to know $r^{\pi\pi}_{PT}$ to extract the value of the CP violating
phase from the CP asymmetry in $B\to\pi\pi$. This is the problem of penguin
pollution.

A variety of solutions to this problem have been proposed.
One type of approach is to exploit the fact that the (strong) penguin
contribution to $P_{\pi\pi}$ is pure $\Delta I={1\over2}$, while the tree
contribution to  $T_{\pi\pi}$ contains a piece which is $\Delta I={3\over2}$.
Isospin symmetry allows one to form a relation among the amplitudes
$B^0\to\pi^+\pi^-$, $B^0\to\pi^0\pi^0$, and $B^+\to\pi^+\pi^0$ and another one
for the charge conjugate processes. A simple geometric construction then allows
one to disentangle the unpolluted $\Delta I={3\over2}$ amplitudes, from which
the CP violating phase may be extracted cleanly \cite{Gronau:1990ka}.
The key experimental difficulty is that one must measure accurately the
flavor-tagged rate for $B^0\to\pi^0\pi^0$. Since the final state consists of
only four photons, and the branching fraction is expected to be of
${\cal O}(10^{-6})$, this is very hard.

Second, one might attempt to calculate the penguin matrix elements.
Model-dependent analyses are not adequate for this purpose, since the goal is
the extraction of fundamental parameters. Precise calculations of such matrix
elements from lattice QCD are far in the future, given the necessity for a
treatment allowing for final state interactions, the large energies of the
$\pi$'s, and the need for an unquenched calculation. Recently, new QCD-based
analyses of the $B\to\pi\pi$ matrix elements have been proposed
\cite{BBNS,Keum:2001ph}.

The third type of approach is to use measurements of $B\to K\pi$ decays  to
determine $|P_{K\pi}|$. Once $|P_{K\pi}|$ is known, flavor $SU(3)$ is used  to
relate $|P_{K\pi}|$ to $|P_{\pi\pi}|$.  It may also be possible to relate
$B_d\to \pi^+\pi^-$ to $B_s\to K^+K^-$ (which is expected to be measured at
the Tevatron) using $SU(3)$~\cite{Fleischer:1999pa}. The problem with these
approaches is that some $SU(3)$ breaking corrections remain a source of
irreducible uncertainty.

An alternative is to perform an isospin analysis of the process $B^0\to\rho\pi
\to\pi^+\pi^-\pi^0$
\cite{Lipkin:1991st,Gronau:1991dq,Snyder:1993mx,Quinn:2000by}. Here one must
study the time-dependent asymmetry over the entire Dalitz plot, probing
variously the intermediate states $\rho^\pm\pi^\mp$ and $\rho^0\pi^0$. The
advantage here is that the final states with two $\pi^0$'s need not be
considered. On the other hand, thousands of cleanly reconstructed events would
be needed. It is yet unclear how well this can be done. With integrated
luminosity of 0.5 ab$^{-1}$, it is possible that a meaningful measurement
will be achieved.

\subsection{\boldmath $B\to DK$ and $B\to D^*\pi$}
One method of extracting the CP violating phase $\gamma$ that seems
theoretically clean involves $B\to KD^0({\overline D^0})$ decays
\cite{Gronau:1991dp,Atwood:1996ci}. The relevant quark transitions are
$b\to c\bar us$ and $b\to\bar cus$. The method
requires high statistics and is probably realistic with 10 ab$^{-1}$.

The final state $D^{*+}\pi^-$ is common to $B^0$ and $\overline{B^0}$ decays.
Therefore, the decay rates are sensitive to interference effects between the
direct decay and the first-mix-then-decay paths, which in turn have interesting
dependence on CP violating phases \cite{Aleksan:1991nh}. By measuring the four
time-dependent decay rates, $B^0,\overline{B^0}\to D^{*\pm}\pi^{\mp}$ one
can determine the amplitude ratio and the CP violating phase $2\beta + \gamma$
in a theoretically clean way.

The problem in this measurement is that
\beq\label{supdec}
{A(\bar b\to\bar uc\bar d)\over A(\bar b\to\bar cu\bar d)}\sim
{V_{ub}^*V_{cd}\over V_{cb}^*V_{ud}}\sim\lambda^2.
\eeq
Thus the interference effects are small and hard to measure.  Despite this
difficulty, its theoretical cleanliness makes this a potentially very important
measurement. With integrated luminosity of 0.5 ab$^{-1}$ the error of
$\sin(2\beta+\gamma)$ is expected to be $0.15-0.20$~\cite{babar36,belleDpi}. A
similar measurement of the four time dependent rates $B_s, \overline{B}_s \to
D_s^\pm K^\mp$ measures $2\beta_s + \gamma$.  The advantage of this mode is
that, unlike Eq.~(\ref{supdec}), the two amplitudes are comparable in size.
The disadvantage is that because the large Cabibbo allowed $B_s \to D_s \pi$
background must be suppressed, this measurement will probably only be doable at
LHCB/BTeV.

An interesting, related proposal was made in ref. \cite{Falk:2001pd}. The
idea is to measure the angle $\gamma$ using CP tagged decays, $B_{\rm
CP}\to DK_S$. Such a measurement can only be done in a very high
luminosity $e^+e^-$ $B$ factory. If such a future collider also operates
at the $\Upsilon(5S)$ resonance, it may be possible to cleanly determine
$\gamma$ from time-integrated measurements of CP tagged $B_s\to D_s K$
decays as well \cite{Falk:2000ga}.

\subsection{\boldmath $B_s\to \psi\phi$ and $\psi\eta^{(\prime)}$}
The CP asymmetry in these processes is the analog of $a_{\psi K_s}$ and
measures the relative phase between $B_s$ mixing and $b\to c\bar c s$ decay.
This is $\sin2\beta_s$, which is presently constrained in the SM to be between
0.026 and 0.048
\cite{Laplace:2002ik}; a larger asymmetry would be a clear sign of new physics.
The expected error at CDF is about $1.6$ times that of $\sin2\beta$, further
diluted by one minus twice the $CP$-odd fraction the $\psi\phi$ final state.
Although this $CP$-odd contribution is expected to be small, it can be avoided
by using the decay modes $B_s\to\psi\eta^{(\prime)}$, which are pure $CP$-even.

\subsection{\boldmath $a_{\phi K_S}$}
Within the Standard Model, where the Kobayashi-Maskawa phase is the only
source of CP violation in meson decays, there are strong correlations
between various CP asymmetries. One of the best known examples is that of the
CP asymmetries in $B\to\psi K_S$ and $B\to\phi K_S$. The two decays proceed
via different quark transitions, $b\to c\bar cs$ for the first, and $b\to s
\bar ss$ for the latter. Yet, the Standard Model predicts that the two CP
asymmetries are equal to within a few percent. This is a result of the fact
that $V_{cb}V_{cs}^*$ and $V_{tb}V_{ts}^*$ are almost aligned. Within
extensions of the Standard Model, the correlation can often be lost due to
significant new contributions to the $b\to s\bar ss$ transition. For example,
in supersymmetric models there could be squark-gluino penguin diagrams that
compete with the SM quark$-W$ diagrams. The comparison of the two CP
asymmetries can therefore cleanly signal new physics \cite{Grossman:1997ke}.

While $a_{\psi K_S}$ has already been measured, this is not the case for
$a_{\phi K_S}$. The problem here is the small branching ratio
\cite{Lista:2001wm},
\beq\label{bpsiks}
{\cal B}(B\to\phi K_S)=(8.1^{+3.1}_{-2.5}\pm0.8)\times10^{-6}.
\eeq
With 0.5 ab$^{-1}$, the accuracy in the CP asymmetry is expected to be $\delta
a_{\phi K_S}\sim0.25$~\cite{babar36}.  The theoretical uncertainty in $a_{\phi
K_S} \simeq a_{\psi K_S}$ is of order $\lambda^2 \sim 0.04$.  Measuring
$a_{\phi K_S}$ with such a small uncertainty requires $\sim
20\,\mbox{ab}^{-1}$, or about two years at a $10^{36}$ cm$^{-2}$ sec$^{-1}$
$e^+ e^-$ machine. This measurement may also be done at hadron colliders, but
no detailed study is available.

This test of new physics would benefit from measurements of $B^+\to\phi\pi^+$
and $B^+\to K^*K^+$ which will constrain rescattering effects
\cite{Grossman:1998gr}.

Before there is sufficient data for measuring $a_{\phi K_S}$, comparison of
$a_{\psi K_S}$ with the asymmetry measured in $b\to c \bar cd$ modes, such as
$B\to D^{(*)} D^{(*)}$, is also interesting. The sensitivity to new physics
contributions to the decay amplitude is, however, smaller.

\subsection{\boldmath $A_{\rm SL}$}
CP violation in $B$ mixing can be measured, for example, in semileptonic
decays: $A_{\rm SL} = [\Gamma(\overline{B}{}^0(t)\to \ell^+) - \Gamma(B^0(t)
\to \ell^-)] / [\Gamma(\overline{B}{}^0(t)\to \ell^+) + \Gamma(B^0(t) \to
\ell^-)]$ is proportional to the relative phase between $\Gamma_{12}$ and
$M_{12}$, analogous to $\mbox{Re}\,(\varepsilon_K)$ in the kaon sector. The SM
predicts the relative phase to be small, suppressed by $m_c^2/m_W^2$
in $B_d$ mixing, and by an additional factor of $\lambda^2$ in $B_s$ mixing.
The best present constraint, based on $\sim20\,\mbox{fb}^{-1}$ data, is $A_{\rm
SL} = (0.48 \pm 1.85)\times 10^{-2}$~\cite{Aubert:2001xc}, while the SM
prediction is at the $10^{-3}$ level \cite{Laplace:2002ik},
\beq\label{aslsm}
-1.3\times10^{-3}< A_{\rm SL}({\rm SM})<-0.5\times10^{-3}.
\eeq
While measuring $A_{\rm SL}$ at the SM level seems impossible even at a very
high luminosity $B$ factory, new physics could significantly enhance the
asymmetry and make it observable. A model independent analysis (with the
only assumption that tree level decays are dominated by SM processes) yields
\cite{Laplace:2002ik}
\beq\label{aslnp}
-0.004< A_{\rm SL}({\rm NP})<+0.04.
\eeq

\subsection{\boldmath $D^0-{\overline{D^0}}$ mixing}
Time dependent measurements of $D\to K\pi$ decays are sensitive to
$D^0-{\overline{D^0}}$ mixing. It is expected that, with an integrated
luminosity of 0.5 ab$^{-1}$, one can be sensitive to the mixing parameters
$x=\Delta m/\Gamma$ and $y=\Delta\Gamma/2\Gamma$ at the level of
$x^2+y^2\sim10^{-5}$. We emphasize that a signal at that level, that is,
$x,y\sim0.003$, may come from the Standard Model long distance contributions
(see, {\it e.g.} \cite{Falk:2001hx}). To disentangle new physics from
Standard Model contributions, it is crucial to measure separately $x$, $y$,
the relevant strong phase and, most important for this purpose, CP violation
\cite{Bergmann:2000id}. Such a program may require a high luminosity
$B$-factory.

\subsection{Rare Decays}
Various rare decays provide useful measurements of the CKM parameters as well
as possible probes of new physics. At present, inclusive rare decays are under
better theoretical control than the exclusive ones, and Table I summarizes some
of the most interesting modes. A clean theoretical interpretation of the latter
requires that we know the corresponding form factors. (Note, however, that CP
asymmetries are independent of the form  factors.) While useful relations
between various form factors can be derived  from heavy quark symmetry,
ultimately unquenched lattice calculations will be  needed for a clean
theoretical interpretation of exclusive decays.  Rare $b\to s$ and $b\to d$
decays are sensitive in the SM to $|V_{td}V_{tb}|$ and $|V_{ts}V_{tb}|$,
respectively.  Thus the $b\to d$ rates are expected to be about a factor of
$|V_{td}/V_{ts}|^2 \sim \lambda^2$ smaller than their $b\to s$ counterparts. As
a guesstimate, in $b\to q\, l_1 l_2$ decays one expects $10-20\%$ $K^*/\rho$
and $5-10\%$ $K/\pi$.

In our introduction we highlighted the fact that with the measurement of
$a_{\psi K}$ the KM mechanism of CP violation has passed its first precision
test. We should emphasize that in the last year we have been learning that the
CKM contributions to rare decays are also likely to be the dominant ones.
Support to this statement comes, for example, from the measurement of ${\cal
B}(B\to X_s\gamma)$ which agrees with the SM at the 15\%
level~\cite{Chen:2001fj}, the measurement of $B\to K\ell^+\ell^-$ which is in
the ballpark of the SM expectation~\cite{Abe:2001dh,Aubert:2001xt}, and the
non-observation of direct CP violation in $b\to s\gamma$ at the 0.2
level~\cite{Coan:2000pu,Aubert:2001me}.  These new results make it less likely
that we will observe orders-of-magnitude enhancement of rare $B$ decays. It is
more likely that only precision measurements and a broad program will be able
to find signals of new physics.

\begin{table}[t]
\begin{tabular}{ccc|cccc}
Decay  &  Approximate  &  Present  &
  \multicolumn{4}{c}{Number of events assuming SM rates}  \\
mode  &  SM rate  &  status  &  $0.5\,\mbox{ab}^{-1}$
  &  $10\,\mbox{ab}^{-1}$  &  CDF/D0  &  BTeV/LHCB  \\ \hline\hline
$B\to X_s\gamma$  &  $3.5\times 10^{-4}$  &  $(3.2 \pm 0.5)\times 10^{-4}$
  &  11K  &  220K  & \\
$B\to K^*\gamma$  &  
  &  $(4.2 \pm 0.5)\times 10^{-5}$
  &  6K  &  120K  &  170  &  25K\\
$B\to X_s\nu\bar\nu$  &  $4\times 10^{-5}$  &  $<7.7\times10^{-4}$
  &  8  &  160  & \\
$B\to \tau\nu$  &  $4\times 10^{-5}$  &  $<5.7\times10^{-4}$
  &  17  &  350  & \\[2pt]
$\matrix{B\to X_s e^+ e^- \cr B\to X_s\mu^+\mu^-}$
  &  $\matrix{8\times 10^{-6} \cr 6\times 10^{-6}}$
  &  $\matrix{<1.0\times10^{-5} \cr <1.9\times10^{-5}}$
  &  300  &  6K  &  &  few K  \\[10pt]
$\matrix{B\to K^*\ell^+\ell^- \cr B\to K\ell^+\ell^-}$
  &  
  &  $\matrix{<2.5\times 10^{-6} \cr (7.5\pm2.7)\times 10^{-7} }$
  &  $100$  &  2K  &  $100$  &  4K \\[6pt]
$B_s\to \tau^+\tau^-$  &  $1\times 10^{-6}$  &  \\
$B\to X_s\tau^+\tau^-$  &  $5\times 10^{-7}$  &   \\
$B\to \mu\nu$  &  $2\times 10^{-7}$  &  $<6.5\times10^{-6}$
  &  8  &  150  & \\
$B_s\to \mu^+\mu^-$  &  $4\times 10^{-9}$  &  $<2\times10^{-6}$
  &  &  &  $<10^{-8}\,^\ddag$  &  $10$ \\
$B\to \mu^+\mu^-$  &  $1\times 10^{-10}$  &  $<2.8\times10^{-7}$
  &  &  &  $<3.5\times 10^{-9}\,^\ddag$  &  $<\,$few \\
\end{tabular}\vspace*{4pt}
\caption{Rare decays~\protect\cite{tabref}.  Future estimates should be taken
with some caution.  The CDF/D0 column corresponds to $2\,\mbox{fb}^{-1}$, the
LHCB/BTeV one to 1 year of running.
$^\ddag$\,Expected upper bounds.}
\end{table}


$(i)$ $B\to X_{s,d}\gamma$ or $B\to K^*\gamma$: provide strong limits
on $m_{H^\pm}$ in 2HDM models, and constrain various supersymmetric models.
The CP asymmetry provides additional constraints on new phyics.  The best
limits are $-0.27 < A_{CP}(B\to X_s\gamma) < 0.10$~\cite{Coan:2000pu} and
$-0.17 < A_{CP}(B\to K^*\gamma) < 0.08$~\cite{Aubert:2001me} at the 90\%\,CL.
In the former case the SM prediction is firmly below 0.01.  The photon
spectrum, which is not sensitive to new physics, is important for
determinations of $|V_{ub}|$ and the $b$ quark mass.

$(ii)$ $B\to X_{s,d}\ell^+\ell^-$ or $B\to K^{(*)}\ell^+\ell^-$: sensitive
probes of new physics that modifies the $bsZ$ coupling.  Of particular interest
are the forward-backward asymmetry, the forward-backward CP asymmetry and the
CP asymmetry in the rate.  Remarkably, the location where the forward-backward
asymmetry in $B\to K^*\ell^+\ell^-$ vanishes (near $q^2 = 4\,\mbox{GeV}^2$ in
the SM) also provides model independent information on short distance
parameters \cite{Burdman:1998mk}.

$(iii)$ $B\to X_{s,d}\nu\bar\nu$ or $B\to K^{(*)}\nu\bar\nu$: probe new
physics that modifies the $bsZ$ coupling, or contain unconstrained couplings
between three 3rd generation fermions \cite{Grossman:1996gt}. This mode is
particularly clean (in some sense the $B$ physics analog of
$K\to\pi\nu\bar\nu$) but experimentally very challenging.

$(iv)$ $B\to\ell\bar\nu$: measures $f_B|V_{ub}|$ in the SM and is sensitive
to new physics, such as charged Higgs.

$(v)$ There are many additional useful probes of new physics, such as
direct CP violation, lepton number or lepton flavor violation,
$B\to\ell^+\ell^-$, etc.


\section{New Physics and Future B Factories}
There are three important goals that can be achieved with future
$B$-factories and will make flavor violating and/or CP violating
processes excellent probes of new physics:

(i) {\it Better precision} in the measurements of processes related, for
example, to the determination of $|V_{ub}|$ or of $|V_{td}/V_{ts}|$, will
allow to observe small deviations from the Standard Model predictions.

(ii) {\it Higher statistics} will allow the measurement of (or improved
upper bounds on) rare decays.

Within the framework of supersymmetry, the improvement in precision and
sensitivity means that we will be able to explore stronger levels of
universality.  Models without universality, such as alignment models, may
give effects of ${\cal O}(\lambda)$ and are already being probed by
present measurements. Models of approximate universality, such as
dilaton dominance (or $U(2)$ models for the first two generations), may
induce effects of ${\cal O}(\alpha_s/\pi)$ (or ${\cal O}(\lambda^2)$),
and will begin to be probed when the experimental and theoretical accuracy
reaches the few percent level. Models of exact universality, such as GMSB,
give only small and calculable deviations and in large parts of their parameter
space give predictions that are similar to the SM.

(iii) A {\it Broad program} will allow to learn detailed features of new
physics and consequently to distinguish between various
models within each class.

For example, the information from $D-\bar D$ mixing and $B-\bar B$ mixing
probes whether the effects of new physics are restricted to either of the up
or down sectors. The comparison of $a_{\phi K_S}$ and $a_{\psi K_S}$ probes
whether the effects of new physics are restricted to either of the $\Delta B=1$
and $\Delta B=2$ processes. The information from $K\to\pi\nu\bar\nu$ decays
can be confronted with that from $B$ factories to teach us whether the effects
are largest for the third generation or significant also for the first two
generations. The information from electric dipole moments can be added to that
from $B$ factories to reveal whether new CP violation is flavor diagonal or
flavor changing or both.

A new era in the study of flavor physics and CP violation has just begun.
The coming decade is expected to significantly enrich our understanding
of these aspects and hopefully teach us about new physics.

\begin{acknowledgments}

Y.N. Thanks the organizers for their hospitality. We thank Adam Falk for
helpful discussions and Michael Dine and Yael Shadmi for comments on the
manuscript.
Z.L.\ was supported in part by the Director, Office of Science, Office of High
Energy and Nuclear Physics, Division of High Energy Physics, of the U.S.\
Department of Energy under Contract DE-AC03-76SF00098.
Y.N.\ is supported by the Israel Science Foundation founded by the
Israel Academy of Sciences and Humanities.

\end{acknowledgments}


\end{document}